**Compression of phase-only holograms with JPEG standard and deep learning**

Shuming Jiao[1,3,4], Zhi Jin[1,4], Chenliang Chang[2], Changyuan Zhou[1], Wenbin Zou[1,*], Xia Li[1]

1. Shenzhen Key Lab of Advanced Telecommunication and Information Processing, College of Information Engineering, Shenzhen University, Shenzhen, Guangdong, China
2. Jiangsu Key Laboratory for Opto-Electronic Technology, School of Physics and Technology, Nanjing Normal University, WenYuan Road 1, 210023, Nanjing, China
3. Tsinghua Berkeley Shenzhen Institute (TBSI), Shenzhen, 518000, China
4. These authors contribute equally to this work
Corresponding author: wzouszu@sina.com
For technical enquiries, please contact albertjiaoee@126.com

Abstract:
It is a critical issue to reduce the enormous amount of data in the processing, storage and transmission of a hologram in digital format. In photograph compression, the JPEG standard is commonly supported by almost every system and device. It will be favorable if JPEG standard is applicable to hologram compression, with advantages of universal compatibility. However, the reconstructed image from a JPEG compressed hologram suffers from severe quality degradation since some high frequency features in the hologram will be lost during the compression process. In this work, we employ a deep convolutional neural network to reduce the artifacts in a JPEG compressed hologram. Simulation and experimental results reveal that our proposed "JPEG + deep learning" hologram compression scheme can achieve satisfactory reconstruction results for a computer-generated phase-only hologram after compression.

1. Introduction

Computer Generated Holography (CGH) allows the recording and reconstruction of a desired complex light wavefront including both amplitude and phase information. With the development of computer and optical technologies, CGH has been extensively applied in many fields such as three-dimensional dynamic holographic display [1-3], holographic projection [4-7], virtual and augmented reality [8,9], optical tweezers [10] and optical information security [11,12]. The research on CGH generation, conversion and compression algorithms receives much attention in the past decade. For example, the enormous amount of calculation in the generation of a complex hologram from a 3D object model consisting of many points is a huge challenge and the fast CGH calculation problem has been investigated from different perspectives [13-17]. Another major concern is that holographic display devices such as spatial light modulator (SLM) usually cannot display both the amplitude and phase part of a complex hologram simultaneously. Fast and high-quality phase-only hologram calculation from an object image is favorable for commonly used phase-only type SLMs. A phase-only hologram (Figure 1) can be calculated from an object image with various methods such as Gerchberg-Saxton iterative algorithm [18-20], error diffusion algorithm [21-24], random phase mask method [25-27] and down-sampling mask method [28-29]. Compared with other methods, error diffusion method [21-24] has several advantages including no forward and backward field propagation, reduction of high frequency noise in the result and direct converting a complex hologram into a phase only one [8]. In this paper, the error diffusion method is employed for the computer generation of phase-only holograms from object images.

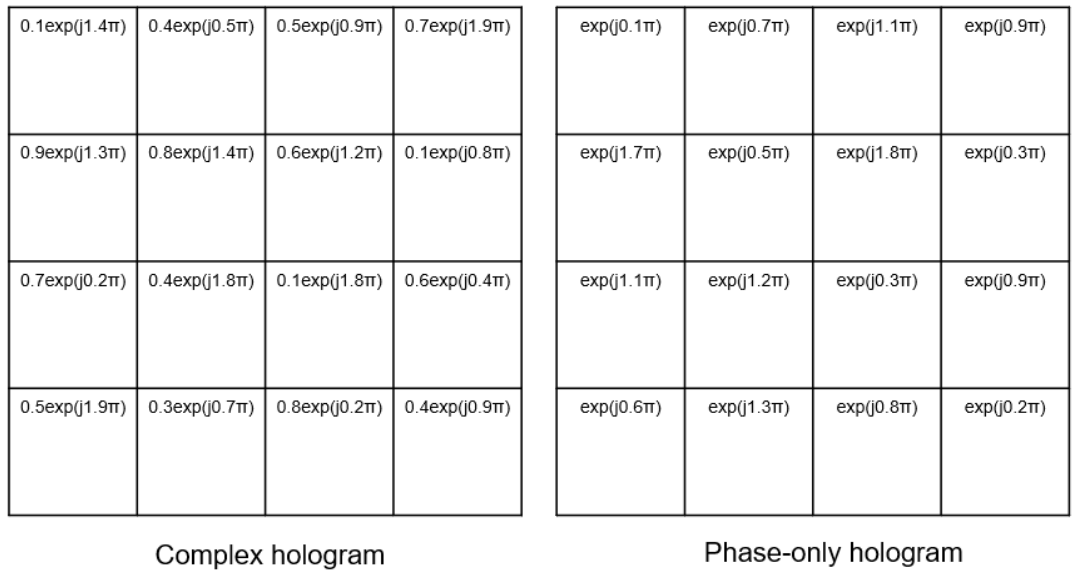

Figure 1 An example of complex hologram (left) and phase-only hologram (right) with a size of 4×4 pixels

In addition to the numerical generation of complex or phase-only holograms, the compression of hologram data is a critical issue. Since the unit pixel size of a hologram is typically rather small (a few micrometers), the entire hologram may contain a huge number of pixels and demands considerable storage cost and transmission bandwidth. It is necessary to develop efficient hologram compression algorithms to reduce the size of data representing a hologram. In the past decade, much research effort has been made on hologram compression [30-34] and a literature review can be found in [30]. In essence, a hologram can be regarded as a special kind of two-dimensional (2D) image. For natural 2D images (e.g. photographs), standardized image compression algorithms such as JPEG [35] have been widely employed. However, some image properties of holograms significantly differ from photographs. For example, a photograph is usually locally smooth while a hologram consists of many high-frequency fringe patterns. As a consequence, JPEG compression often cannot give optimal compression performance for holograms [30]. Instead of directly adopting JPEG standard, many customized compression algorithms are proposed for holograms such as Fresnelets [31], Wavelet-Bandelets Transform

[32], vector quantization [33] and enhanced wavelet transform [34]. These customized hologram image compression methods can yield very satisfactory performance for certain type of holograms. However, on the downside, these customized algorithms are not compatible with commonly used JPEG standard, which is supported very extensively by almost every computer system and digital device while these customized compression methods are not very generally adopted. A hologram compression method is very favorable if it has compatibility with JPEG standard and can achieve good compression efficiency at the same time. In this work, we proposed a phase-only hologram compression scheme incorporating JPEG standard and deep convolutional network post-processing.

This article is organized as follows. In Section 2, the error diffusion method for phase-only hologram generation is briefly described. In Section 3, the working principles of JPEG image compression standard are discussed and our proposed artifact reduction scheme for JPEG compressed holograms using deep convolutional network is presented. In Section 4, simulation and experimental results are demonstrated to verify the effectiveness of our proposed scheme. In Section 5, a brief conclusion is provided.

## 2. Computer generated phase-only hologram with error diffusion method

Error diffusion algorithm [21-24] is a non-iterative and high-quality method for calculating a phase-only hologram from an object image on computer. The working principles of phase-only hologram calculation with error diffusion method are described below. To start with, $O(x, y)$ denotes the object image and a complex Fresnel hologram $H(x, y)$ can be calculated based on the Fresnel diffraction formulas illustrated by Equation (1) and (2).

$$f(x, y; z) = \frac{\exp(i\frac{2\pi z}{\lambda})}{i\lambda z} \exp[i\frac{\pi(x^2 + y^2)}{\lambda z}] \quad (1)$$

$$H(x, y) = O(x, y) * f(x, y; z) \quad (2)$$

where $f(x, y; z)$ denotes the impulse function for Fresnel transform, $\lambda$ denotes the wavelength of illumination light and $z$ denotes the distance between the object image plane and hologram plane.

The complex hologram $H(x, y)$ can be converted to a phase-only hologram $P(x, y)$ with error diffusion algorithm [21-24] to achieve high-quality holographic display on a phase-only spatial light modulator. In the error diffusion method, the complex value of each hologram pixel is forced to be unity amplitude and the resulting complex error is diffused to neighboring pixel values. The following operations (from Equation (3) to Equation (8)) are performed on each individual holographic pixel sequentially in row-by-row and column-by-column scanning manner, illustrated in Figure 2.

$$E(x_j, y_j) = H(x_j, y_j) - P(x_j, y_j) \quad (3)$$

$$H(x_j, y_j+1) \leftarrow H(x_j, y_j+1) + w_1 E(x_j, y_j) \quad (4)$$

$$H(x_j+1, y_j-1) \leftarrow H(x_j+1, y_j-1) + w_2 E(x_j, y_j) \quad (5)$$

$$H(x_j+1, y_j) \leftarrow H(x_j+1, y_j) + w_3 E(x_j, y_j) \quad (6)$$

$$H(x_j+1, y_j+1) \leftarrow H(x_j+1, y_j+1) + w_4 E(x_j, y_j) \quad (7)$$

$$P(x_j, y_j) = Phase[H(x_j, y_j)] \quad (8)$$

where $P(x_j, y_j)$ denotes the phase-only pixel value at position $(x_j, y_j)$ after the complex pixel value $H(x_j, y_j)$ is phase truncated (the magnitude is forced to be unity), $E(x_j, y_j)$ denotes the complex error and four different weighting coefficients $w_1=7/16$, $w_2=3/16$, $w_3=5/16$ and $w_4=1/16$ are imposed on four different directions, shown in Figure 2.

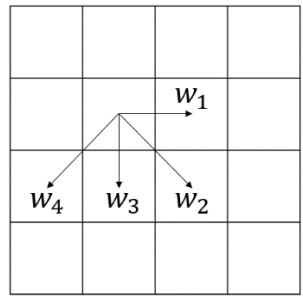

Figure 2. Propagation of errors to four different neighboring pixels with corresponding weighting coefficients in error diffusion algorithm

After each pixel in a complex hologram is processed, the entire hologram will become a phase-only type. More details about error diffusion methods for phase-only hologram generation can be found in [21-24]. In practice, bi-directional error diffusion, a slightly modified version of the above unidirectional error diffusion, can yield slightly better performance [21].

## 3. JPEG image compression and proposed artifact reduction scheme by deep convolutional network

Each pixel in a phase-only hologram has an intensity value ranging from 0 to $2\pi$ and a phase-only hologram can be regarded as a gray-scale intensity image. Various image compression algorithms can be attempted for the compression of phase-only holograms.

JPEG is a very commonly used lossy compression scheme for digital images, which was firstly proposed in 1992 by Joint Photographic Experts Group [35]. In JPEG scheme, the original image (e.g. a gray-level photograph or a phase-only hologram) is first divided into 8×8 pixel blocks and each individual block undergoes discrete cosine transform (DCT). Then the coefficients in the transformed domain of each block are quantized with more quantization levels for low frequency components and less quantization levels for high frequency components. Subsequently, zig-zag coding, entropy coding and Huffman coding are performed on the quantized coefficients. Finally, the original image becomes a compressed bit stream with much smaller data size than the original uncompressed image. A reconstructed image with certain image degradation can be obtained when the JPEG binary bit stream is decompressed with the inverse procedures as compression.

JPEG compression can significantly reduce the data size of an original natural image and only introduce minor quality degradation in the decompressed image. The reason is that a natural image (like a photograph) is usually locally smooth and its most information is concentrated in the low frequency coefficients of DCT spectrum. The discarded high frequency information in the quantization step has negligible effect on the image quality. However, phase-only holograms have substantially different image characteristics compared with photographs [36]. One feature of a hologram is that it contains many high frequency fringe patterns. If JPEG compression is directly applied to a phase-only hologram, the decompressed result after compression will suffer from heavy damage and the holographically reconstructed image from the JPEG compressed hologram will be severely degraded as well.

In this work, we employ deep convolutional neural networks to reduce the artifacts of JPEG compressed phase-only holograms. In the past few years, deep learning methods receive much attention and exert considerable impact in many fields. Very recently, deep learning has been introduced to holographic research area and gained success in different applications [37-42].

The network structure employed in our task is illustrated in Figure 3(a), based on the previous works of restoring JPEG compressed photographs [43]. The overall flowchart of our proposed "JPEG + deep learning" hologram compression scheme is illustrated in Figure 3(b). The network contains four convolutional layers, feature extraction layer, feature enhancement layer, non-linear mapping layer and reconstruction layer. The network input and output are both image patches with a size of $31 \times 31$ pixels extracted from a hologram. The input hologram is processed patch by patch and the restored output patches are stitched to constitute the restored hologram. The artifacts caused by JPEG compression in phase-only holograms such as high frequency noise and blocking effect can be suppressed by the deep convolutional network.

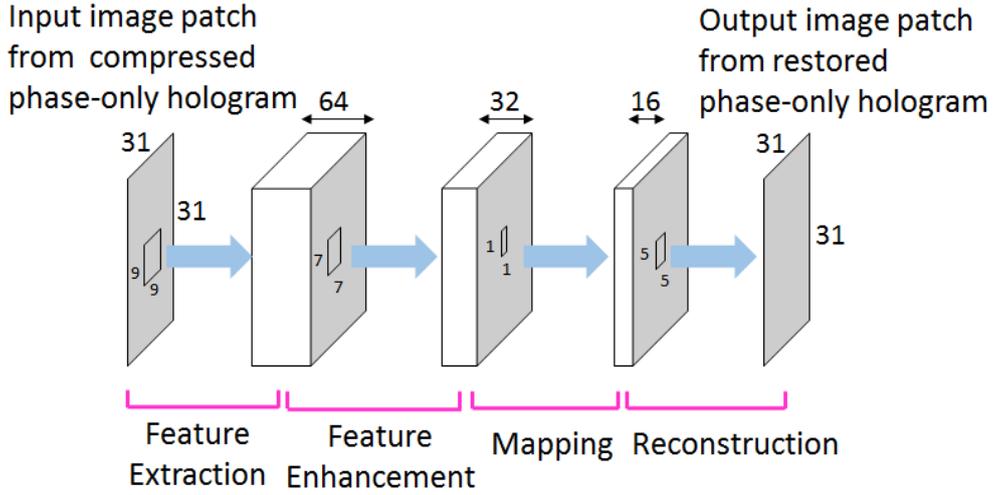

(a)

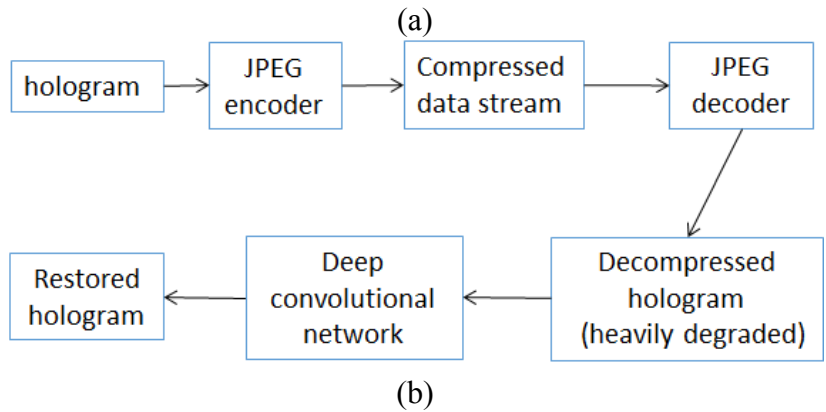

(b)

Figure 3. (a) Structure of deep convolutional network for the quality enhancement of JPEG compressed holograms in our work; (b) Overall flowchart of our proposed "JPEG+deep learning" hologram compression scheme.

In the training, the patches from JPEG compressed hologram are employed as the input of the network and the patches from intact uncompressed hologram are employed as the target output (ground truth). The mean square error (MSE) between the ground truth and the network output at each iteration is minimized by adjusting the network weighting coefficients with stochastic gradient descent backpropagation. The network weighting parameters are trained separately for phase-only holograms generated at different distances.

## 4. Results and discussion

In our work, ten pairs of compressed and uncompressed phase-only holograms (1024×1024 pixel size) calculated from ten different object images (512×512 pixel size) with error diffusion method are employed as the training holograms. Two compressed phase-only holograms generated from 'Pepper' and 'Pirate' images respectively are employed as the test holograms. The number of training holograms in this work is much less than conventional deep learning works because different holograms usually have many similar image blocks and fringe patterns. The replication properties in hologram images allows a small number of examples to contain most universal hologram features, which has been revealed in previous works [44]. The ten training holograms are divided into patches (31×31 pixels) and the patches are augmented by rotation and mirroring. A total number of 960384 hologram patches are fed into the network for training. The object images for generating training holograms and testing holograms are shown in Figure 4 and Figure 5.

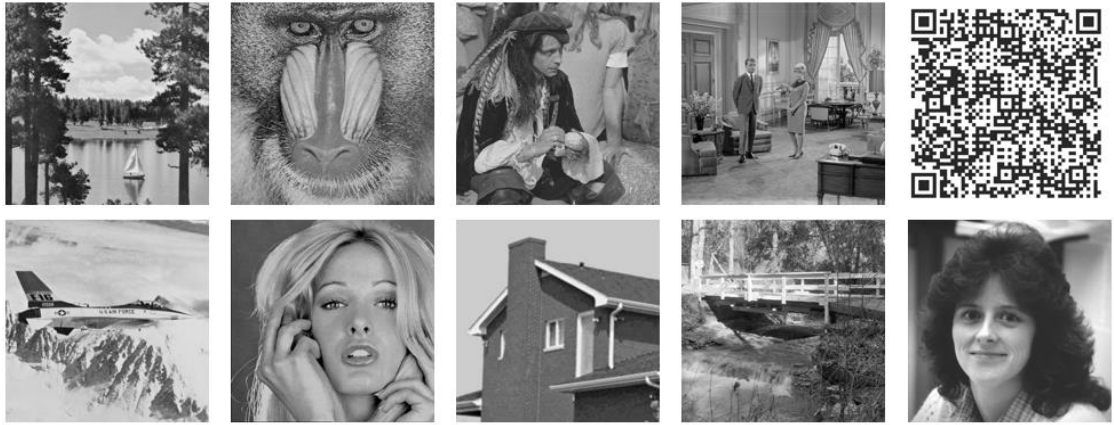

Figure 4. Ten object images employed for generating training holograms

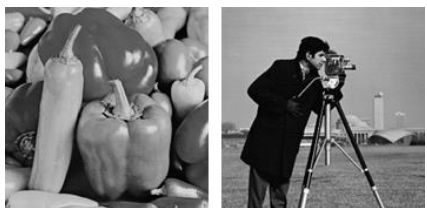

Figure 5. Two object images ("Pepper" and "Cameraman") employed for generating testing holograms

In the computer calculation of phase-only holograms, the optical wavelength of illumination light is set to be 532nm, the pixel size is 8μm and the distance between object image and hologram plane is 0.3m or 0.5m. Both the ten training holograms and two testing holograms are then compressed by JPEG when the quality factor is set to 1. The JPEG compressed hologram is restored by our proposed deep convolutional network. One example of the original hologram, JPEG compressed hologram and restored hologram with our proposed scheme is shown in Figure 6. It can be observed that some high frequency patterns are deteriorated in the hologram after JPEG compression and restored to certain extent after the quality enhancement using our proposed network.

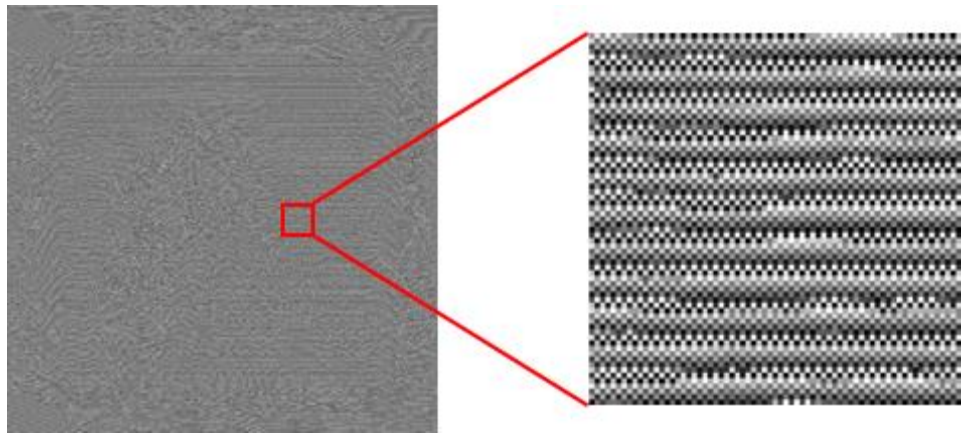

(a)

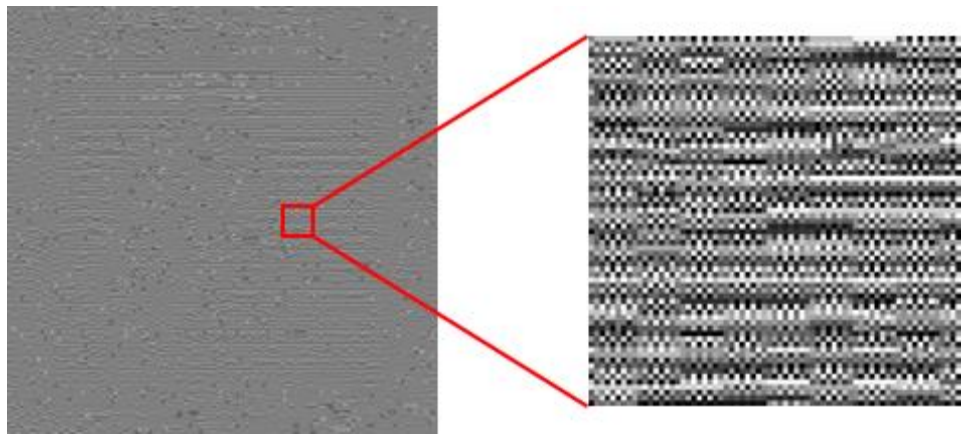

(b)

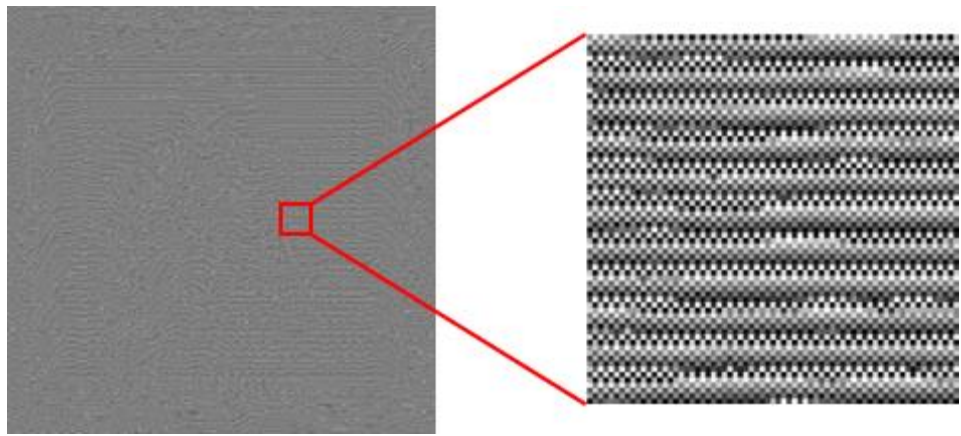

(c)

Figure 6. (a)Original hologram; (b)JPEG compressed hologram; (c)Restored hologram with our proposed scheme.

The reconstructed images from original uncompressed holograms, JPEG compressed holograms and restored holograms are compared in numerical simulations. The reconstructed images are shown in Figure 7. The PSNR and SSIM values of the reconstructed results from compressed holograms and restored holograms, in comparison with the reconstructed results from uncompressed holograms, are presented in Table 1. The compression ratio is approximately 7 and the data size of original hologram, 1024 KB (1 MB), is reduced to around 142 KB to 144 KB after JPEG compression. The PSNR and SSIM values reveal that our proposed scheme can significantly enhance the reconstructed image quality from JPEG compressed holograms.

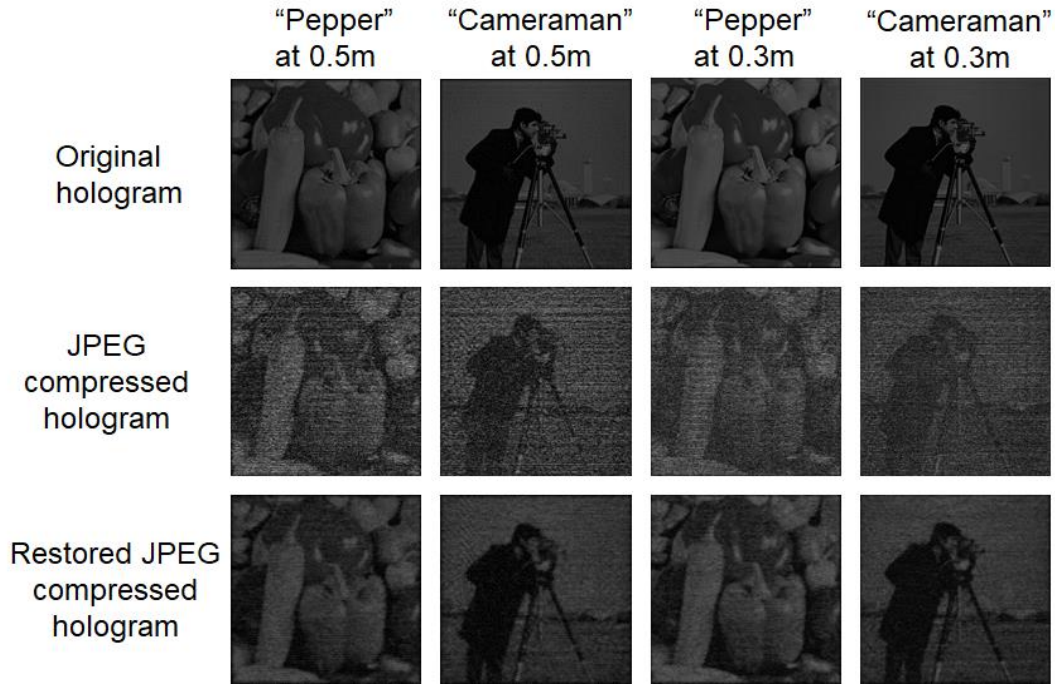

Figure 7. Reconstructed images from original holograms, JPEG compressed holograms and restored holograms in numerical simulation.

Table 1 PSNR and SSIM values of reconstructed results from JPEG compressed holograms and restored holograms with our proposed scheme

|  |  | "Cameraman" image | | "Pepper" image | |
| --- | --- | --- | --- | --- | --- |
|  |  | 0.5m | 0.3m | 0.5m | 0.3m |
| Compression ratio | | 7.2113 | 7.1111 | 7.2113 | 7.1608 |
| Reconstructed image from compressed hologram | PSNR | 19.10 dB | 17.83 dB | 18.92 dB | 17.64 dB |
| | SSIM | 0.1651 | 0.0967 | 0.2007 | 0.1036 |
| Reconstructed image from restored hologram | PSNR | 28.86 dB | 26.86 dB | 29.88 dB | 27.16 dB |
| | SSIM | 0.6036 | 0.4465 | 0.6767 | 0.4852 |

Our proposed scheme is further verified in optical experiments and the calculated holograms are optically reconstructed using an optical setup [6] shown in Figure 8. The hologram is reconstructed with a phase-only Holoeye PLUTO spatial light modulator (SLM). The system parameters are the same as the numerical simulation (wavelength: 532nm; pixel size: 8μm; hologram size: $1024 \times 1024$ pixels;

object-hologram distance: 0.3m/0.5m). The phase-only holograms loaded into the SLM are assigned with an additional carrier phase in order to transversally shift the desired reconstructions away from the un-diffracted (zero-order) beam component. The projected beam from the SLM is transmitted through a 4-f optical filtering system containing a low-pass iris with 3.3mm diameter to block the zero-order noise. The optically reconstructed results are shown in Figure 9. It can be observed that the quality of reconstructed images after hologram enhancement with deep learning are improved, compared with the ones from compressed holograms without restoration.

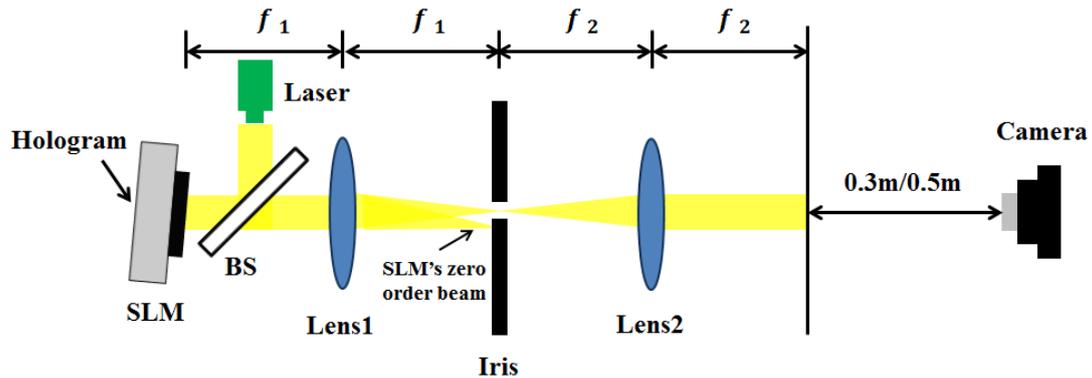

Figure 8. Optical setup for hologram reconstruction experiment.

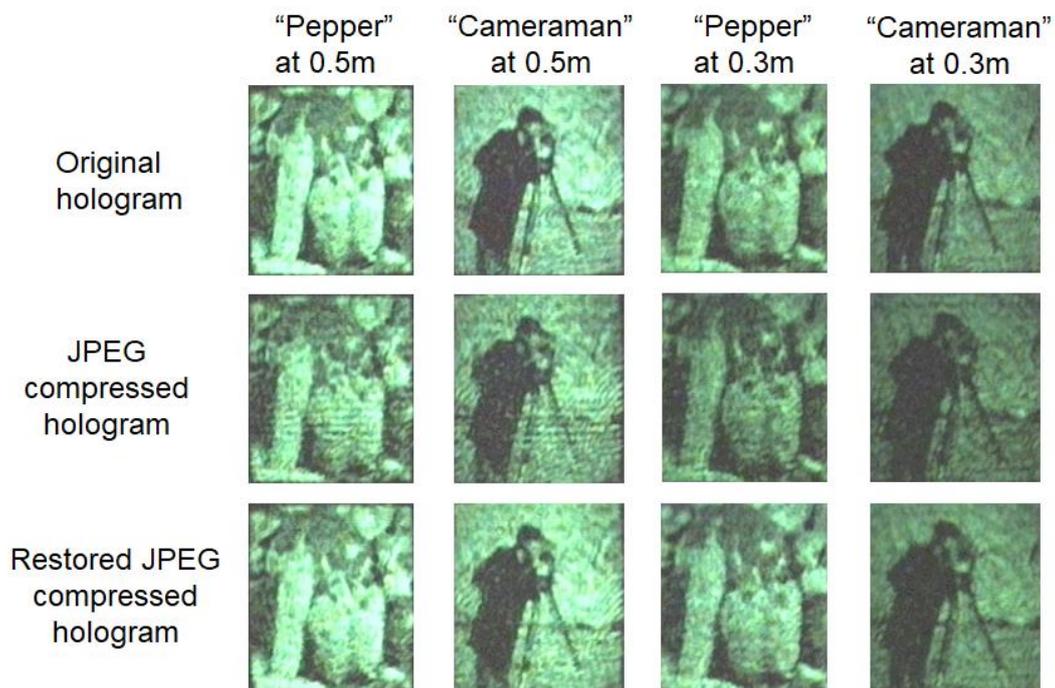

Figure 9. Reconstructed images from original holograms, JPEG compressed holograms and restored holograms in optical experiments.

## 5. Conclusion

It is essential to reduce the enormous amount of data representing a computer-generated phase-only hologram in the processing, transmission and storage. JPEG photograph compression standard can be attempted for hologram compression with an advantage of universal compatibility, compared with customized hologram compression algorithms. Deep convolutional networks can be employed to reduce the artifacts for a JPEG compressed hologram. Simulation and experimental results reveal that our proposed "JPEG + deep learning" hologram compression scheme can maintain the reconstructed image quality when the data size of original hologram is significantly reduced.